\documentclass[12pt]{article}

\def\be{\begin{equation}}
\def\ee{\end{equation}}
\usepackage{euscript,amsfonts,amssymb,epsfig,latexsym}

\begin{document}
\title{Interacting Stochastic Process and Renormalization Theory}
\author{Yaroslav Volovich\thanks{E-mail: \tt yaroslav\_v@mail.ru}\\
~\\
{\it Physics Department, Moscow State University}\\
{\it Vorobievi Gori, 119899 Moscow, Russia}}
\date{~}
\maketitle
\begin{abstract}

A stochastic process with self-interaction as a model of quantum field
theory is studied. We consider an Ornstein-Uhlenbeck stochastic process
$x(t)$ with interaction of the form $x^{(\alpha)}(t)^4$, where $\alpha$
indicates the fractional derivative. Using Bogoliubov's $R-$operation we
investigate ultraviolet divergencies for the various parameters $\alpha$.
Ultraviolet properties of this one-dimensional model in the case
$\alpha=3/4$ are similar to those in the $\varphi^4_4$ theory but there are
extra counterterms. It is shown that the model is two-loops renormalizable.
For $5/8 \leq \alpha < 3/4$ the model has a finite number of
divergent Feynman diagrams. In the case $\alpha=2/3$ the model is similar to
the $\varphi^4_3$ theory. If $0 \leq \alpha < 5/8$ then the model does not
have ultraviolet divergencies at all. Finally if $\alpha > 3/4$ then the
model is nonrenormalizable.

This model can be used for a non-perturbative study of ultraviolet
divergencies in quantum field theory and also in  theory of phase
transitions.

\end{abstract}

\section{Introduction}

There is a very fruitful interrelation between probability theory and
quantum field theory \cite{BS}-\cite{AccLuVol}. In this note  we consider a
stochastic process that shows the same divergencies as quantum
electrodynamics or $\varphi^4$ theory in the 4-dimensional spacetime.
Therefore this one-dimensional model can be used for studying the fundamental
problem of non perturbative investigation of renormalized quantum field
theory \cite{BS,GJ}. It can also find applications in theory of phase
transitions \cite{KogWil,Sin,PatPok}.

The simplest nontrivial model of quantum field theory in $d-$dimensional
spacetime is the model of scalar field with the $\varphi^4$-interaction. In
one dimensional case ($d=1$) this model is equivalent to an anharmonic
oscillator and it does not have ultraviolet divergencies. In this note
we will study a more interesting model with the interaction that contains
fractional derivatives. It is well known that trajectories of the Wiener
process are H\"older-continuous with exponent $\alpha<1/2$. The property of
a function to be H\"older-continuous is related to the property of having a
fractional derivative.

Let us remind that the free scalar {\it massless} quantum field on the
semi-axis is the Wiener process. The scalar {\it massive} one-dimensional
quantum field is an Ornstein-Uhlenbeck stochastic process $x(t)=x(t,\omega)$
\cite{H}. In order to introduce an {\it interacting} Ornstein-Uhlenbeck
stochastic process we will use fractional derivatives $x^{(\alpha)}(t)$ ($0
\leq \alpha < 1$) \cite{Vla,FC}. Stochastic differential equations with
fractional derivatives \cite{Vla2x} are considered in \cite{BikVol} on
$p-$adic number fields.

In this note an interacting Ornstein-Uhlenbeck stochastic process with
interaction of the form $x^{(\alpha)}(t)^4$ will be discussed. Using
Bogoliubov's $R-$opera\-tion we investigate ultraviolet divergencies for
various parameters $\alpha$. Ultraviolet properties of this one-dimensional
model in the case $\alpha=3/4$ are similar to those in the $\varphi^4_4$
theory but there are extra counterterms. It is shown that the model is
two-loops renormalizable. For $5/8 \leq \alpha < 3/4$ the model has a
finite number of divergent Feynman diagrams. In the case $\alpha=2/3$ the
model is similar to the $\varphi^4_3$ theory. If $0 \leq \alpha < 5/8$ then
the model does not have ultraviolet divergencies at all. Finally if
$\alpha > 3/4$ then the model is nonrenormalizable.

This paper is organized as follows. In the next section the interacting
stochastic process is defined. Feynman diagrams for this process are
described in Sect.\ref{isp}. A brief discussion of $R-$operation is given in
Sect.\ref{roper}. Analysis of divergent Feynman diagrams for the interacting
stochastic process is presented in Sect.\ref{divdg}. Finally, in
Sect.\ref{examp} interesting particular cases of the interacting stochastic
process are discussed.

\section{The Interacting Stochastic Process}
\label{isp}

Let $x(t)=x(t,\omega)$ be an Ornstein-Uhlenbeck stochastic process with the
correlation function
\be
\label{propagator}
E(x(t)x(\tau))=F(t-\tau)
=\frac{1}{2\pi}\int^{\infty}_{-\infty} \frac{e^{ip(t-\tau)}}{p^2+m^2}dp
=\frac{e^{-m|t-\tau|}}{2m}
\ee
where $m>0$. There exists a spectral representation of the
Ornstein-Uhlen\-beck stochastic process \cite{GihSko}
$$
x(t,\omega)=\int e^{ikt} \zeta(dk,\omega)
$$
where $\zeta(dk,\omega)$ is a stochastic measure. We define the fractional
derivative as
\be
\label{eq01x}
x^{(\alpha)}(t,\omega)=\int |k|^{\alpha} e^{ikt} \zeta(dk,\omega)
\ee
If $0\leq\alpha <1/2$ then $x^{(\alpha)}(t)$ is a stochastic process. If
$\alpha\geq 1/2$ then one needs a regularization described below. We will
use distribution notations and write
$$
\zeta(dk,\omega)=\tilde{x}(k,\omega)dk
$$
$$
\tilde{x}(k,\omega)=\frac{1}{2\pi}
\int^{\infty}_{-\infty} x(t,\omega) e^{-ikt} dt
$$

We want to give a meaning to the following correlation functions
\be
\label{corrf}
K(t_1,\ldots,t_N)=\left.
E(x(t_1) \cdots x(t_N) e^{-\lambda U}) \right/ E(e^{-\lambda U})
\ee
for all $N=1,2,\ldots$ Here
\be
\label{U-def}
U=\int^{\infty}_{-\infty} x^{(\alpha)}(\tau)^4 g(\tau) d\tau
\ee
where $g(\tau)$ is a nonnegative test function with a compact support (the
volume cut-off), $x^{(\alpha)}(t)$ denotes the fractional derivative
(\ref{eq01x}) and $\lambda \geq 0$. We will denote the expectation value as
$E(A)=\left< A \right>$. In this notations
$$
\left< x(t)x(\tau) \right>
=\frac{1}{2\pi}\int^{\infty}_{-\infty} \frac{e^{ip(t-\tau)}}{p^2+m^2}dp
$$

If $\alpha\geq 5/8$ then the expectation value in (\ref{corrf}) has no
meaning even if we expand it into the perturbation series in $\lambda$
because there are ultraviolet divergencies (see below). We have to introduce
a cutoff stochastic process $x_{\varkappa}(t)$ \cite{GJ}
$$
x_{\varkappa}(t,\omega)=\int^{\varkappa}_{-\varkappa}e^{ikt}\zeta(dk,\omega)
$$
Instead of $U$ in (\ref{corrf}) we put
$$
U_{\varkappa}=\int~:x_{\varkappa}^{(\alpha)}(\tau)^4:~g(\tau) d\tau
$$
where
$$
x^{(\alpha)}_{\varkappa}(t,\omega)=
\int^{\varkappa}_{-\varkappa} |p|^{\alpha} \tilde{x}(p,\omega) e^{ipt} dp
$$
Here
$$
:x_{_\varkappa}^{(\alpha)}(\tau)^4:=
\int\limits^{\varkappa}_{-\varkappa}
\cdots\int\limits^{\varkappa}_{-\varkappa}
dp_1\cdots dp_4 |p_1|^{\alpha}\cdots |p_4|^{\alpha}
e^{i\tau (\sum^4_{j=1}p_j)}
$$
$$
:\tilde{x}(p_1)\tilde{x}(p_2)\tilde{x}(p_3)\tilde{x}(p_4):
$$
where the normal product is defined by the relation (Wick's theorem)
\be
\label{pair}
:\tilde{x}(p_1)\tilde{x}(p_2)\tilde{x}(p_3)\tilde{x}(p_4):=
\tilde{x}(p_1)\tilde{x}(p_2)\tilde{x}(p_3)\tilde{x}(p_4)
\ee
$$
-\left< \tilde{x}(p_1)\tilde{x}(p_2) \right>
:\tilde{x}(p_3)\tilde{x}(p_4):+\cdots
$$
$$
-\left< \tilde{x}(p_1)\tilde{x}(p_2) \right>
\left< \tilde{x}(p_3)\tilde{x}(p_4) \right>
-\cdots
-\left< \tilde{x}(p_1)\tilde{x}(p_4) \right>
\left< \tilde{x}(p_2)\tilde{x}(p_3) \right>
$$
and
$$
\left< \tilde{x}(p)\tilde{x}(k) \right>
=\frac{1}{2\pi}\cdot\frac{\delta(p+k)}{p^2+m^2}
$$

The problem is to prove that after the renormalization there exists a limit
of the correlation functions
$$
\left<x(t_1) \cdots x(t_N) e^{-\lambda U_{\varkappa}}\right>_{ren}
$$
as $\varkappa\to\infty$. We will consider this problem below by using the
Bogoliubov-Parasiuk $R$-operation.

{\bf Remark}. Quadratic interaction
$$
U=\int x^{(\alpha)}(\tau)^2 g(\tau) d\tau
$$
is a well defined stochastic variable if $\alpha < 1/2$. One can go beyond
the boundary $\alpha=1/2$ if one takes the normal product
$$
U=\int~:x^{(\alpha)}(\tau)^2:~g(\tau) d\tau
$$

An heuristic definition of the correlation function (\ref{corrf}) is given
by the functional integral
\be
\label{corf1}
K(t_1,\ldots,t_N)=
\left.\int \varphi(t_1)\cdots\varphi(t_N) e^{-S} {\EuScript D}\varphi\right/
\int e^{-S} {\EuScript D}\varphi
\ee
where the action
\be
\label{action}
S=\int \left[ \frac{1}{2}\dot{\varphi}(t)^2+\frac{m^2}{2}\varphi(t)^2+
\lambda g(t)\varphi^{(\alpha)}(t)^4 \right] dt
\ee
and $\varphi(t)$ is a real valued function on the real axis.

\section{Feynman Diagrams}

Feynman diagrams (or graphs) are convenient tools for labeling and recording
terms that appear while integrating polynomials with respect to a Gaussian
measure, see \cite{BS,GJ}. For the Ornstein-Uhlenbeck stochastic process one
has
\be
\label{zvezda}
\left<x(t_1)\ldots x(t_l)\right>
=\sum_{pairings} F(t_{i_1}-t_{i_2})\cdots F(t_{i_{l-1}}-t_{i_l})
\ee
where $F(t)$ is defined by (\ref{propagator}) and the sum extends over all
distinct ways of choosing the pairs $\{t_{i_k}, t_{i_{k+1}}\}$.

For the correlation function (\ref{corrf}) one has the perturbative
expansion
$$
\left<x(t_1)\ldots x(t_N) e^{-\lambda U}\right>
=\sum_{n=0}^{\infty} \frac{(-\lambda)^n}{n!}
\left<x(t_1)\cdots x(t_N) U^n\right>
$$
Then by using (\ref{zvezda}) one writes
$$
\left<x(t_1)\cdots x(t_N) U^n\right>=
$$
$$
=\int d\tau_1 \cdots d\tau_N
\left<
x(t_1)\ldots x(t_N)x^{(\alpha)}(\tau_1)^4 \cdots x^{(\alpha)}(\tau_n)^4
\right>=
$$
$$
=\sum_{\{\Gamma\}} K_{\Gamma} (t_1,\ldots,t_N)
$$
where $\Gamma$ ranges over a set of graphs and $K_{\Gamma} (t_1,\ldots,t_N)$ is a
function of $t_1,\ldots,t_N$ assigned to $\Gamma$. A graph (or diagram) is a
collection of vertices (represented as points), lines
(represented as line segments joining vertices), and legs (represented as
line segments which attach at one endpoint only to a vertex).

First let us describe a Feynman graph that corresponds to
$$
A=x(t_1)\ldots x(t_N)x^{(\alpha)}(\tau_1)^4 \cdots x^{(\alpha)}(\tau_n)^4
$$
Each factor $x(t_i)$ is represented by a vertex with a leg and each factor
$x^{(\alpha)}(\tau_i)^4$ in $A$ is represented by a single vertex with $4$
legs. Now the integral
$$
\int d\tau_1\cdots d\tau_N \left<A\right>
$$
is represented by a sum of graphs obtained by pairing the legs in the graph
corresponding to $A$ in all possible manners. Examples of Feynman diagrams
are presented in (Fig.\ref{fig:fish}-\ref{fig:eight}).

Now each variable $t_j$ is represented by a leg. Each variable $\tau_i$ is
represented by a vertex with $4$ legs or lines. Each line joining vertices
$s_k$ and $s_l$ (here $s_k$ stands for $t_j$ or $\tau_i$) gives rise to a
function $F(s_i-s_k)$ defined in (\ref{propagator}). One has
$$
K_{\Gamma} (t_1,\ldots,t_N)
=\int d\tau_1 \cdots d\tau_N
\prod_{\mathop{\scriptstyle lines}\limits_{\scriptstyle legs}} F(s_k-s_l)
$$

In momentum representation we obtain the expression of the form
$$
\left<\tilde{x}(p_1)\ldots\tilde{x}(p_N)e^{-\lambda U}\right>
=\sum_{\{\Gamma\}} G_{\Gamma}(p_1,\ldots,p_N)
$$

The sum runs over all diagrams $\Gamma$ with $N$ external legs that can
be build up using $4$-vertices corresponding to the $x^{(\alpha)4}$ term.
Contributions from the connected diagrams with $n$ $4-$vertices come from
the expression
\newpage
$$
\int\left< \tilde{x}(\right.p_1)\ldots\tilde{x}(p_N)
(\prod^{4}_{i=1}\tilde{x}(k^{(1)}_i)|k^{(1)}_i|^{\alpha} )
\ldots ( \prod^4_{i=1}\tilde{x}(k^{(n)}_i)|k^{(n)}_i|^{\alpha}\left.)
\right>
$$
$$
\delta (\sum^4_{i=1} k^{(1)}_i)\cdots \delta (\sum^4_{i=1} k^{(n)}_i)
\prod_{i,j} dk^{(j)}_i
$$
Using the pairing (\ref{pair}) we obtain the contribution from a diagram
$\Gamma$ with $L$ internal lines in the form
\be
\label{form1}
\int (\prod^L_{l=1} \frac{|k_l|^{2\alpha}}{k^2_l + m^2})
(\prod^n_{i=1} \delta (\xi_i)) dk_1\cdots dk_L
\ee
Here $\xi_i$ denotes the linear combination of momenta assigned
to the $i$-th vertex. We integrate over $(n-1)$ momenta using
$\delta(\xi_i)$ and obtain
\be
\label{form2}
\delta (\sum^N_{j=1} p_j) \int I_{\Gamma} dk_1\cdots dk_l
\ee
where $l=L-(n-1)$.
Here $I_{\Gamma}$ is the unrenormalized integrand of the form
\be
\label{ig}
I_{\Gamma}=\prod^L_{j=1} \frac{|q_j|^{2\alpha}}{q^2_j+m^2}
\ee
where $q_l$ are linear combinations of the internal momenta
$k_1,\ldots ,k_L$ and external momenta $p_1,\ldots ,p_N$.

{\bf Remark}. If we make a change of variables $x^{(\alpha)}(t)=y(t)$ in the
original Lagrangian then we obtain an interacting stochastic process with
$\lambda y(t)^4$ interaction and propagator
$$
\left< \tilde{y}(p)\tilde{y}(k) \right>
=\frac{1}{2\pi}\cdot\frac{|p|^{2\alpha}\delta(p+k)}{p^2+m^2}
$$
This remark explains the appearance of the factor $|q_j|^{2\alpha}$ in
(\ref{ig}).

\section{$R$-Operation}
\label{roper}

A diagram $\Gamma$ is called proper (or one-particle-irreducible) if it is
connected and can not be separated in two parts by cutting a single line.
The canonical degree $D(\Gamma)$ of a proper diagram is defined by the
dimension of the corresponding Feynman integral with respect to the
integration variables. From (\ref{form1}) and (\ref{form2}) we have
\be
\label{dgree}
D=D(\Gamma)=(2\alpha -2)L+l=(2\alpha-1)L-n+1
\ee
A diagram $\Gamma$ is called superficially divergent if its dimension
$D(\Gamma) \geq 0$. A proper diagram $\Gamma$ which is superficially
divergent ($D(\Gamma) \geq 0$) is called renormalization part. A
$\Gamma$-forest $W$ is a set of diagrams satisfying the following
conditions:
\begin{enumerate}
\item the elements of $W$ are renormalization parts of $\Gamma$
\item any two elements of $W$ are non-overlapping
\end{enumerate}
Let $I_{\Gamma}$ be an unrenormalized integrand. The Bogoliubov-Parasiuk
prescription for the renormalized integral can be written as
\cite{BS,Zim}
$$
R_{\Gamma}=\sum_{W} \prod_{\gamma \in W} (-t^{\gamma}) I_{\Gamma}
$$
where the sum runs over all $\Gamma$-forests. Here $t^{\gamma}I_{\Gamma}$
denotes the Taylor series with respect to the appropriate external variables
around zero up to order $D(\gamma)$. The renormalized integral $R_{\Gamma}$
is considered as a result of action of $R$-operation to $I_{\Gamma}$.
This formula is equivalent to the original recursive $R$-operation defined
by using the reduced diagrams. It is one of the main results of the
Bogoliubov-Parasiuk-Hepp-Zimmerman (BPHZ) theory that one has the absolute
convergence of the renormalized integral
$$
J_{\Gamma}=\int dk_1\cdots dk_l R_{\Gamma}
$$
The same result can be obtained if we introduce the cut-off $\varkappa$, then
make a reparametrization of the parameters in the original (bare)
Lagrangian, and finally remove the cut-off. For example the renormalized
Euclidean interaction Lagrangian of the $\lambda\varphi^4$ theory in
$4-$dimensional space is \cite{BS}
$$
{\EuScript L}=\frac{Z_3-1}{2}\left(
(\triangledown\varphi)^2+\frac{m^2}{2}\varphi^2
\right)+\frac{\delta m^2}{2}\varphi^2+\lambda(Z_4-1)\varphi^4
$$
Here $Z_3$, $Z_4$, and $\delta m^2$ are represented as series in $\lambda$
and the corresponding coefficients in the expansions of $Z_3$ and $Z_4$
have logarithmic divergencies as $\varkappa\to\infty$ and coefficients of
$\delta m^2$ have quadratic divergencies.

We will show at the two-loop level that in the $x^{(\alpha)4}$ model with
$\alpha=3/4$ the renormalized Lagrangian has a similar structure
$$
{\EuScript L}=\frac{Z_3-1}{2}\left(
\dot{x}(t)^2+\frac{m^2}{2}x(t)^2
\right)+\frac{\delta m^2}{2}x(t)^2
$$
$$
+\frac{Z_2-1}{2}x^{(3/4)}(t)^2+\lambda(Z_4-1)x^{(\alpha)}(t)^4
$$
Note the term $x^{(3/4)}(t)^2$ which is discussed below.

\section{Divergent Diagrams}
\label{divdg}

The action (\ref{action}) and the correlation functions (\ref{corrf}),
(\ref{corf1}) describe a one dimensional model of quantum field theory.
Although one-dimensional it is an interesting and instructive model because
it demonstrates the most striking property of quantum field theory, i.e.
its ultraviolet divergencies. By expanding (\ref{corrf}) or (\ref{corf1})
into the perturbation series in $\lambda$ one obtains the Feynman diagrams
(\ref{form1}). Let us investigate the convergence of the analytical
expressions corresponding to the Feynman diagrams by using the canonical
degree $D(\Gamma)$ of the diagram (\ref{dgree})
\be
\label{eq11x}
D=L(2\alpha-1)-n+1
\ee
If for a given diagram $D < 0$ then it is superficially
finite, otherwise it is divergent.

Let us consider a proper diagram with $n$ vertices, $L$ internal
lines, and $E$ legs. We have the following relation
\be
\label{eq10x}
4n=2L+E
\ee
Note that for any nontrivial connected diagram
\be
\label{eq12x}
2n \geq L \geq n \geq 2
\ee
\be
\label{eq12'}
E \leq 2n
\ee

\noindent{\bf Theorem}~~{\it If $\alpha < 5/8$ then all Feynman diagrams of
the interacting stochastic process are superficially finite. If $5/8 \leq
\alpha < 3/4$ then there exists a finite number of divergent diagrams,
moreover all divergent diagrams have only $0$ or $2$ legs. If $\alpha=3/4$
then the model is renormalizable and all divergent diagrams have only $0$,
$2$  or $4$ external lines. Finally, if $\alpha > 3/4$ then the model is
nonrenormalizable.}

\noindent{\bf Proof}~~Let us prove the first statement of the theorem, i.e.
if $\alpha < 5/8$ then $D<0$ for any $n\geq 2$. Using (\ref{eq11x}) and
(\ref{eq12x}) we have
\be
\label{eq13x}
D\Big|_{\alpha < 5/8} < 2L\cdot\frac{5}{8}-L-n+1=\frac{L-4n+4}{4}\leq
\ee
$$
\leq \frac{2n-4n+4}{4}=\frac{2-n}{2} \leq 0
$$
From (\ref{eq13x}) it follows that $D<0$ for any $\alpha < 5/8$.

Let us consider $\alpha=5/8$. Similarly to (\ref{eq13x}) from (\ref{eq11x})
we have
\be
\label{eq14x}
D\Big|_{\alpha=5/8}=\frac{L-4n+4}{4}\leq\frac{2-n}{2}\leq 0
\ee
Therefore only two-point ($n=2$) diagram could be
divergent (in this case $D=0$). Rewriting (\ref{eq14x}) in the form
\be
\label{eq14y}
D\Big|_{\alpha < 5/8}=\frac{4-(E+L)}{4}
\ee
From (\ref{eq14y}) it follows that only diagram with
$E\!=\!0,~L\!=\!4,~n\!=\!2$ (Fig. \ref{fig:nut}) is divergent.

In the case when $5/8 < \alpha < 3/4$ we can write
\be
\label{eq16x}
\alpha=\frac{3}{4}-\varepsilon
\ee
where $0<\varepsilon<1/8$. Substituting
(\ref{eq16x}) into (\ref{eq11x}) and using (\ref{eq12x}) we have
\be
\label{eq17x}
D\Big|_{\alpha=3/4-\varepsilon}=\frac{L}{2}-2L\varepsilon-n+1
\leq\frac{2n}{2}-2n\varepsilon-n+1=1-2n\varepsilon
\ee
Thus for any given $\varepsilon>0$ (and therefore
any $\alpha<3/4$) there exists a number $N$ such that for any $n\geq N$ the
canonical dimension $D<0$. Hence there exists only a finite number of divergent
diagrams. Rewriting (\ref{eq17x}) in the form
$$
D\Big|_{\alpha=3/4-\varepsilon}=-2L\varepsilon+\frac{4-E}{4}
$$
It follows that $D\geq 0$ only if $E<4$, i.e. $E=0$ or $E=2$ and the model is
super-renormalizable.

Let us consider the case when $\alpha=3/4$. Using
(\ref{eq10x}) and (\ref{eq11x}) we have
\be
\label{eq18x}
D\Big|_{\alpha=3/4}=1-\frac{E}{4}
\ee
The equality (\ref{eq18x}) means that all divergent diagrams have only
$0,~2,~\mbox{or}~4$ legs and the model is renormalizable.

Finally if $\alpha>3/4$ we have
\be
\label{eq19x}
D\Big|_{\alpha>3/4}=\frac{L}{2}-n+1=\frac{2n-E+1}{2}\geq\frac{1}{2}>0
\ee
Therefore if $\alpha>3/4$ then all proper diagrams are divergent. $\Box$

\section{Examples}
\label{examp}

\subsection{The stochastic process with $\alpha=3/4$ and $\varphi^4_4$
theory}

In this section the case $\alpha=3/4$ is discussed in more detail. The
stochastic process with this $\alpha$ is very interesting because the
structure of ultraviolet divergencies in this case is similar to that of the
$\varphi^4_4$ theory. Indeed, all proper vacuum diagrams as well as $2-$ and
$4-$leg diagrams are divergent.

However let us note an important difference. For the two-loop self-energy
diagram (Fig. \ref{fig:sun}) we have
$$
\Sigma_{\varkappa}(p)=\int^{\varkappa}_{-\varkappa} dk
\int^{\varkappa}_{-\varkappa} dq
\frac{|k|^{3/2}|q|^{3/2}|k+q-p|^{3/2}}
{(k^2+m^2)(q^2+m^2)((k+q-p)^2+m^2)}
$$
The leading divergence of this integral as $\varkappa\to\infty$ is
proportional to $\varkappa^{1/2}$. Therefore the renormalized Lagrangian
should have a counter-term of the form
$$
C\varkappa^{1/2}~:x^{(3/4)}(t)^2:
$$
The fractional derivative $x^{(3/4)}$ appears in this expression due to a
factor $|p|^{\alpha}$ that according to the Feynman rules discussed in
Sect.\ref{roper} corresponds to every external leg.

The thorough consideration of the stochastic process with $\alpha=3/4$
requires a further work.

\subsection{The stochastic process with $\alpha=5/8$}

The stochastic process with $\alpha=5/8$ is the simplest stochastic process
that has ultraviolet divergencies. From Theorem it follows that in this
case only the vacuum diagram (Fig. \ref{fig:nut}) is divergent.
The contribution of this diagram is proportional to
$$
\int\limits^{\varkappa}_{-\varkappa}dk_1
\int\limits^{\varkappa}_{-\varkappa}dk_2
\int\limits^{\varkappa}_{-\varkappa}dk_3
\frac{|k_1|^{5/4} |k_2|^{5/4} |k_3|^{5/4} |k_1+k_2+k_3|^{5/4}}
{(k^2_1+m^2)(k^2_2+m^2)(k^2_3+m^2)((k_1+k_2+k_3)^2+m^2)}
$$

\subsection{The stochastic process with $\alpha=2/3$ and $\varphi^4_3$
theory}

\begin{figure}
 \begin{minipage}[b]{.50\linewidth}
  \begin{center}
   \epsfig{file=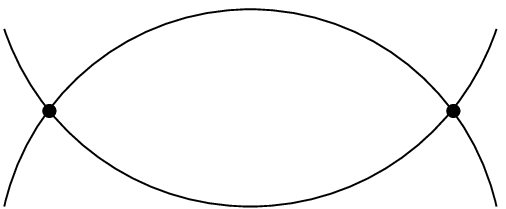,width=.7\linewidth,height=2cm}
   \caption{}\label{fig:fish}
  \end{center}
 \end{minipage}\hfill
 \begin{minipage}[b]{.50\linewidth}
  \begin{center}
   \epsfig{file=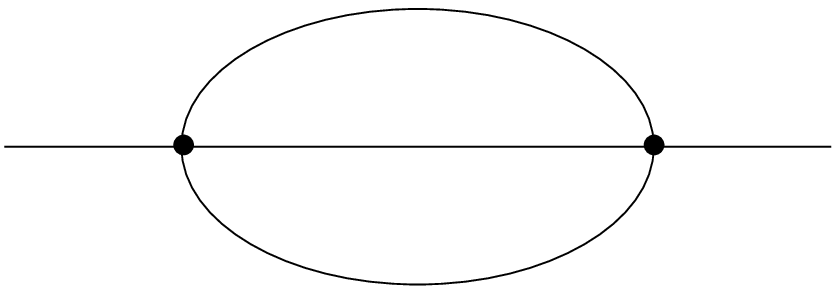,width=.7\linewidth,height=2cm}
   \caption{}\label{fig:sun}
  \end{center}
 \end{minipage}
 \begin{minipage}[b]{.50\linewidth}
 \vspace*{1cm}
  \begin{center}
   \epsfig{file=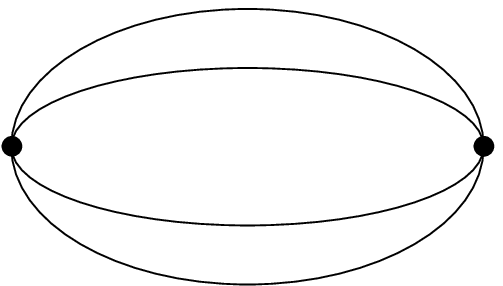,width=.5\linewidth,height=2cm}
   \caption{}\label{fig:nut}
  \end{center}
 \end{minipage}\hfill
 \begin{minipage}[b]{.50\linewidth}
  \begin{center}
   \epsfig{file=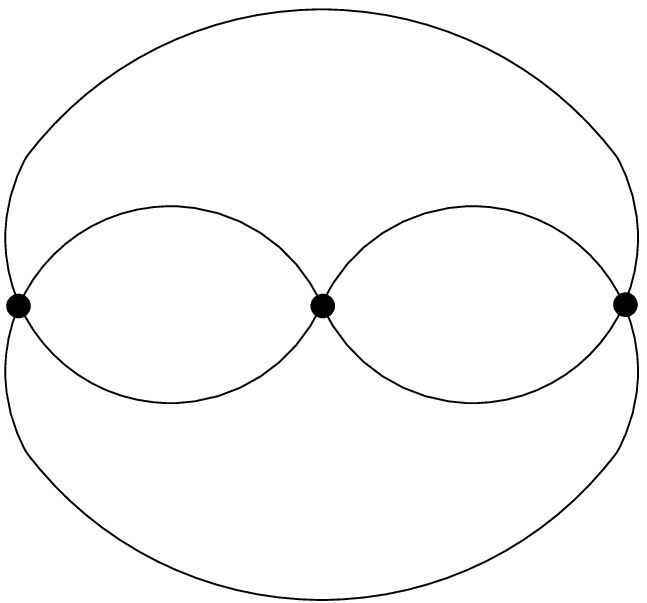,width=.5\linewidth,height=2cm}
   \caption{}\label{fig:eight}
  \end{center}
 \end{minipage}
\end{figure}

The interacting stochastic process with $\alpha=2/3$ is interesting because
it is the simplest stochastic process which has non-vacuum ultraviolet
divergencies. From Theorem it follows that in this case only diagrams shown
on (Fig. \ref{fig:sun},\ref{fig:nut},\ref{fig:eight}) are divergent. The
structure of divergencies is similar to the $\varphi^4_3$ theory
\cite{GliJaf2}, however there is a difference. Besides the divergent
constants corresponding to the vacuum diagrams the renormalized Lagrangian
should also contain the term
$$
C\ln{\varkappa}~:x^{(2/3)}(t)^2:
$$
Here the divergence comes from the two-loop integral
$$
\Sigma_{\varkappa}(p)=\int^{\varkappa}_{-\varkappa} dk
\int^{\varkappa}_{-\varkappa} dq
\frac{|k|^{4/3}|q|^{4/3}|k+q-p|^{4/3}}
{(k^2+m^2)(q^2+m^2)((k+q-p)^2+m^2)}
$$

Renormalized correlation functions are
$$
\left.\left< x(t_1) \cdots x(t_N) e^{-\lambda U_{\varkappa}}\right>
\right/ \left<e^{-\lambda U_{\varkappa}}\right>
$$
where
$$
U_{\varkappa}=\int\,:x_{\varkappa}^{(2/3)}(\tau)^4:\,g(\tau) d\tau
+C\lambda\ln\varkappa \int\,:x_{\varkappa}^{(2/3)}(\tau)^2:\,g(\tau) d\tau
$$
We do not write the contribution from the vacuum diagrams.

\section{Discussions and Conclusions}

In this paper we have considered renormalization of the interacting
stochastic process that has ultraviolet divergencies similar to those that
appear in multidimensional models of quantum field theory. We have studied
the stochastic process in perturbation theory. It would be interesting to
establish the existence of the stochastic process for $\alpha=2/3$
nonperturbatively by using methods used in $\varphi^4_3$ theory
\cite{GliJaf2}. Then properties of trajectories could be studied.

Especially interesting is the case when $\alpha=3/4$ which is similar to the
$\varphi^4_4$ theory. In this case the proof of the renormalizability of the
stochastic process even in perturbation theory requires a further studying.
We have discussed the renormalization in this case only at two-loop level.
It would be also very interesting to investigate the critical behavior of
the interacting stochastic process in the strong coupling regime on the
lattice.

\section{Acknowledgments}

The author is grateful to L. Accardi for invitation to the Centro Vito Volterra
Universit\'a di Roma "Tor Vergata",
to H.-H.Kuo for stimulating discussions of renormalization of
quantum white noise, and
to I.V. Volovich for constant attention to this work.


\end{document}